\documentclass{kluwer}
\usepackage{graphicx}

\newcommand{\citeN}[1]{\citeauthor{#1}\ (\citeyear{#1})}
\newcommand{\citeNP}[1]{\citeauthor{#1},\ \citeyear{#1}}

\newcommand{\apj}{{\it Astrophys.~J.}}

\newcommand{\aap}{{\it Astron.~Astrophys.}}

\newcommand{\solphys}{{\it Solar~Phys.}}

\newcommand{\ujlisti}{
\itemsep=0 em
\parsep=0.5 em
\partopsep=0.25 em
\topsep=0 em}
\newcommand{\ujlistii}{
\itemsep=0 em
\parsep=0.5 em
\partopsep=0.25 em
\topsep=0 cm}

\newenvironment{lista}{\begin{list}{--}{\ujlisti}}{\end{list}}

\renewcommand{\[}{\begin{equation}}
\renewcommand{\]}{\end{equation}}
 
\newcommand{\vc}[1]{\mbox{\bf #1}}

\newcommand{\ov}{\overline}

\begin{document}                                                                                   
\begin{article}
\begin{opening}

\title{THE ROLE OF ACTIVE REGIONS IN THE GENERATION OF TORSIONAL OSCILLATIONS}
\subtitle{}

\author{K. \surname{Petrovay}}
\author{E. \surname{Forg\'acs-Dajka}}
\institute{E\"otv\"os University, Dept.~of Astronomy, Budapest, Pf.~32,
	   H-1518 Hungary} 

\date{[{\it Solar Physics}, {\bf 205}, 39--52 (2002)]}

\runningtitle{The Role of Active Regions in Torsional Oscillations}
\runningauthor{Petrovay \& Forg\'acs-Dajka}

\begin{abstract} 
We present a model for torsional oscillations where the inhibiting effect of
active region magnetic fields on turbulence locally reduces turbulent viscous 
torques, leading to a cycle- and latitude-dependent modulation of the
differential rotation. The observed depth dependence of torsional oscillations
as well as their phase relationship with the sunspot butterfly diagram are
reproduced quite naturally in this model. The resulting oscillation amplitudes
are significantly smaller than observed, though
they depend rather sensitively on model details. Meridional circulation is
found to have only a weak effect on the oscillation pattern.
\end{abstract}


\end{opening}

\section{Introduction}

Solar torsional oscillations, discovered by \citeN{Howard+LaBonte:torso},
consist of alternating latitudinal bands of faster and slower than average
rotation in the solar photosphere, distributed symmetrically to the equator,
and migrating during the solar cycle. Early investigations
(\citeNP{LaBonte+Howard:torso}; \citeNP{Snodgrass+Howard:torso}) revealed a 
characteristic phase relationship between these torsional waves, of wavenumber 
$\sim 2/$hemi\-sphere, and the sunspot butterfly diagram: sun\-spot activity varies
approximately in phase with the latitudinal shear
$\partial\omega/\partial\theta$. The characteristic amplitude of the
oscillations is order of $0.1\,\%$, or a few nHz.

Since the first seismic detection of torsional waves from $f$-mode splittings
by \citeN{Kosovichev+Schou:torso}, helioseismic observations have brought
great advance in the study of these oscillations (\citeNP{Schou+:difrot};
\citeNP{Howe+:deep.torso}). With these methods the depth dependence of these
motions could also be studied. It was found that the motions extend to about
the upper third of the solar convective zone, while coherent torsional
oscillations seem to disappear below a depth of about 70 Mm
(\citeNP{Antia+Basu:torso}; \citeNP{Howe+:SOGO}). These studies also shed light
on the long disputed high latitude behaviour of the oscillations, demonstrating
the existence of a poleward propagating branch at high latitudes
(\citeNP{Howe+:SOGO}). This parallels the existence of a similar branch in the
sunspot/facula butterfly diagram, and it explains earlier propositions of the
coexistence of a $\sim 1/$hemisphere modulation with the torsional waves. 

Most theoretical attempts to interpret the torsional oscillations attribute
them to some feedback effect of the magnetic fields on the differential
rotation. (See, however, \citeNP{Tikhomolov:SOGO} for a dissenting view.)
\citeN{Kichat+:torso} plausibly classify the models as ``macrofeedback'' and
``microfeedback'' scenarios, depending on whether the feedback is due to the
Lorentz force associated with the large-scale magnetic field or to the
inhibiting effect of magnetic fields on the (turbulent) angular momentum 
transport mechanisms responsible for differential rotation. 
The first macrofeedback models were proposed by \citeN{Schussler:torso} and
\citeN{Yoshimura:torso}, while the latest, most extensive and detailed such
models were developed by Covas et~al. (\citeyear{Covas+:torsoletter},
\citeyear{Covas+:torso}). Microfeedback models have been elaborated by the
Potsdam--Irkutsk group (e.g.\ \citeNP{Kuker+:torso}; \citeNP{Kichat+:torso}).

All the models of torsional oscillations proposed to date are based on a
complete mean-field dynamo. At first sight this may seem natural: however, it
is not necessarily a desirable feature of such a model, for two reasons.
Firstly, there is still wide disagreement concerning just how the solar dynamo
operates (cf. review by \citeNP{Petrovay:SOLSPA}). Thus, the use of a
particular dynamo model introduces a high degree of arbitrariness into the
study of torsional oscillations. Second, the recently discovered shallowness of
the  torsional oscillation phenomenon suggests that near-surface magnetic
fields play a dominant role in their generation. These fields are directly
observable in the photosphere, which not only makes it possible to dispense
with the use of a dynamo model to calculate them, but in fact it demonstrates
that a purely  mean-field calculation may perhaps never grasp the most
important factors modulating the angular momentum distribution near the
surface. Indeed, from observations it is well known that the overwhelming
majority of the cycle-dependent part of photospheric magnetic flux density
$\ov{|\vc B|}$ is in the form of active regions (\citeNP{Harvey:PhD},
\citeyear{Harvey:NATO}). The dynamics of the thick active region flux tubes is
governed by volume forces such as buoyancy and the curvature force; it is
therefore largely independent of the external turbulence, and it cannot be
described by one-fluid models like mean field theory, in contrast to the
``passive'' fields consisting of thin fibrils (cf.\ \citeNP{Petrovay:NATO}).

The aim of this paper is to investigate whether active regions (ARs) can play a
role in the modulation of angular momentum distribution in the shallow layers
of the convective zone. An obvious argument against such a role is that the
volume filling factor of the magnetic flux tubes forming ARs is
small throughout the convective zone. This is indeed the case if one regards
the volume actually magnetized; however, the presence of a sufficiently dense
network of individual flux tubes (e.g. a ``magnetic tree'' below a large active
region or an ensemble of ephemeral active regions) may significantly affect
the flow throughout that network, so that the influence of ARs may
extend to a much larger fluid volume. This implies that the rotational
modulation induced by ARs, if any, should not work by the direct
action of Lorentz forces on the large-scale flow but by the spatially more
extended influence on turbulent angular momentum transport. In other words,
AR-induced rotational modulation will be a microfeedback effect.

In a simplified description, this feedback will essentially consist in a
suppression of turbulent viscosity. (Another possible mechanism by which ARs
may modify the differential rotation may be their influence on the large scale
flow.) Indeed, there is now a rather wide consensus that the solar differential
rotation is mainly generated by the $\Lambda$-effect in the deep layers of the
convective zone (possibly aided by the return flow of meridional circulation;
\citeNP{Rekowski+Rudiger}; \citeNP{Gilman:tacho.review}). The pole-to-equator
transport of angular momentum in the deep layers is compensated by a transport
of the opposite sense in the upper half of the convective zone, mainly due to
turbulent viscosity (and possibly to meridional circulation). It is this latter
transport that the active regions field may interfere with, so that in a
simplified approach their influence may be represented by a reduction of the
turbulent viscosity.  This reduction is not expected to extend very deep into
the convective zone as the ``branches'' of the magnetic trees underlying active
regions are thought to separate at a depth of about 50 Mm only.

The structure of our paper is the following. In Section 2 we describe our model
and discuss the problem of the quantitative formulation of the AR-induced
reduction of viscosity. In Section 3 we present the resulting torsional
oscillation profiles as functions of latitude, radius and time, for various
assumptions on the reduction of diffusivity, with and without meridional
circulation. Finally, Section 4 concludes the paper.

\section{The Model}

\subsection{Geometry} 

Our computational volume is a spherical half-shell below the solar surface.
Axial and equatorial symmetry is assumed. The bottom of the shell is placed to
a depth of $z_0=160\,$Mm. It is assumed that the $\Lambda$-effect operates
below our shell only, its effect being described by a prescribed differential
rotation profile at the lower boundary. The bottom depth of 160 Mm may seem too
deep for this assumption to be valid; however, in order to convincingly show
that the torsional waves appearing in our model are confined to depths less
than 70 Mm, the lower boundary had to be placed sufficiently deep below that
level.

\subsection{Equations}

In order to describe differential rotation in the solar interior, it is 
useful to write the equations in a frame rotating with  a fixed
angular velocity $\Omega_0$. The equation of motion reads
\begin{equation}
\partial_t \mathbf{v} + \left( \mathbf{v} \cdot \nabla \right) \mathbf{v} + 
2\vec\Omega_0 \times \mathbf{v} = 
-\nabla V - \frac{1}{\rho} \nabla p + \frac{1}{\rho} \nabla \cdot \mathbf{\tau},
\end{equation}
where $\rho$ and $p$ are density and pressure, respectively, $\vc v$ is the 
fluid  velocity, $V$ is the gravitational potential, and $\tau$ is the viscous
stress  tensor. This equation is supplemented by the constraint of mass
conservation (anelastic approximation):
\begin{equation}
\nabla\cdot ( \rho \mathbf{v}) = 0.
\end{equation}
As a result of the anelastic approximation, the circulation in the spherical 
shell may be represented by a stream function $\Psi$ so that the velocity field 
can be written as
\begin{eqnarray}
v_r &=& \frac{1}{\rho r^2 \sin\theta} \partial_{\theta} \Psi \\
v_{\theta} &=& -\frac{1}{\rho r \sin\theta} \partial_r \Psi \\
v_{\phi} &=& r \sin\theta\, \omega, 
\end{eqnarray}
where $v_r$, $v_{\theta}$ and $v_{\phi}$ are, respectively, the radial,
meridional and  azimuthal components of the velocity, and $\omega$ is the
angular velocity in our rotating frame. In order to present more transparent
equations we write the stream function in the  following form:
\begin{equation}
\Psi = \psi(r) \sin^2\theta \cos\theta                  \label{eq:Psi}
\end{equation}
\begin{equation}
\psi(r) = -0.01\, R_{\odot}^2
         \exp\left[-500\left(\frac{r-r_m}{R_{\odot}}\right)^2\right],
\end{equation}
where $r_m=r_{str}+(R_{\odot}-r_{str})/2$ and $r_{str}=520\,$Mm. The parameters
and form of this formula were designed to mimick the observed properties of
meridional circulation, with a peak amplitude of $\sim 20\,$m/s at the surface
(\citeNP{Komm+:circul}; \citeNP{Latushko:circul}).

\begin{figure}[!t]
\centering
\includegraphics[width=8cm]{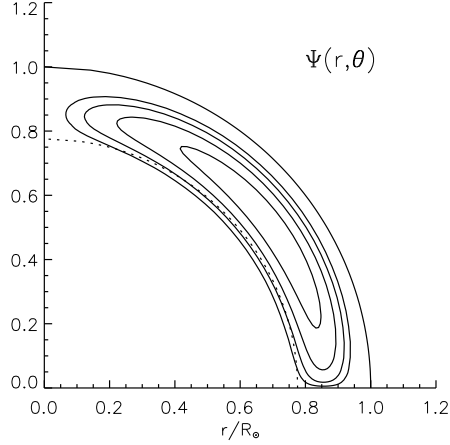}
\caption{Streamlines of the meridional circulation prescribed in
eq.~(\ref{eq:Psi}) (clockwise circulation). The dotted line is the bottom of our
computational domain.}
\label{fig:stream}
\end{figure}

The components of the viscous stress tensor appearing in the azimuthal
component of the Navier-Stokes equation read
\begin{eqnarray}
  \tau_{\theta \phi} &=& \tau_{\phi \theta} = \rho \nu \frac{\sin\theta}{r}\, 
     \partial_\theta \left( \frac{v_{\phi}}{\sin\theta} \right) ,\\ 
  \tau_{\phi r} &=& \tau_{r \phi} =  \rho \nu r \,\partial_r \left( 
     \frac{v_{\phi}}{r} \right),
\end{eqnarray}
where $\nu$ is the viscosity. Thus, the azimuthal component of the equation of 
motion, including the effects of  diffusion, Coriolis force and meridional
circulation, becomes
\begin{eqnarray}
\partial_t \omega &=& \left( \partial_r \nu + 4 \frac{\nu}{r} + 
  \frac{\nu}{\rho} \partial_r \rho
  \right) \partial_r \omega + \nu \partial^2_r \omega \label{eq:main} 
  + \frac{3 \nu \cos\theta}{ r^2 \sin\theta} \partial_{\theta} \omega \\
&+& \frac{\nu}{r^2} \partial^2_{\theta} \omega + M + C \nonumber \\
\nonumber \\ 
M &=& \frac{\psi \omega}{r^3 \rho} \left( 2 \sin^2 \theta - 4 \cos^2 \theta \right) +
   \frac{\psi \partial_r\omega}{r^2 \rho} \left( \sin^2\theta - 2 \cos^2\theta \right) \\
&+& \frac{\partial_r \psi}{r^2 \rho} \left( \partial_{\theta} \omega \cos\theta \sin\theta + 
   2 \omega \cos^2\theta \right) \nonumber \\
C &=& \frac{2 \Omega_0 \psi}{r^3 \rho} \left( \sin^2\theta - 2 \cos^2\theta \right) + 
 \frac{2 \Omega_0 \cos^2\theta \partial_r\psi}{r^2 \rho},
\end{eqnarray}
where $M$ denotes the terms associated with the advection by meridional 
circulation and $C$ denotes the terms associated with the Coriolis force.

\subsection{Boundary and initial conditions}

For the integration of equation (\ref{eq:main}) boundary conditions on
$\omega$ are needed. At the pole and the equator symmetry is required:
\begin{equation}
\partial_\theta \omega =0 \qquad \mbox{at } \theta=0^\circ \mbox { and }
\theta=90^\circ
\end{equation}
At the lower boundary of our computational volume, 
$r_{\mbox{\scriptsize{in}}}=R_\odot-z_0=540\,$Mm,
we suppose that the rotation rate can be described with
the same expression as on the surface. In accordance with the observations of
\citeN{Schou+:difrot}, the  following expression is used for
$\Omega_{\mbox{\scriptsize{surf}}}$:
\begin{equation}
\frac{\Omega_{\mbox{\scriptsize{surf}}}}{2\pi} = 455.4 - 52.4 \cos^2\theta - 81.1 \cos^4\theta \hspace{0.2cm} \mbox{nHz}.
\end{equation}
This is used to give the inner boundary condition on $\omega$,
\begin{equation}
\Omega_0 + \omega = \Omega_{\mbox{\scriptsize{surf}}}
\end{equation}
$\Omega_0$ is (arbitrarily) chosen as the rotation rate in the radiative
interior below the tachocline. On the basis of helioseismic measurements, this
value is equal to the rotation rate of the convection zone at a latitude of
about $30^{\circ}$, corresponding to $\Omega_0/2\pi = 437$ nHz.

Finally, the upper boundary condition on $\omega$, at 
$r_{\mbox{\scriptsize{surf}}}=R_{\odot}$ follows from  the requirement that the 
tangential viscous stress $\tau_{\phi r}$ must vanish at the surface:
\begin{equation}  
\partial_r \omega = 0 \hspace{0.5cm} \mbox{at} 
\hspace{0.1cm}r = r_{\mbox{\scriptsize{surf}}}.
\end{equation}

\noindent
The initial conditions chosen for all calculations are
\begin{eqnarray}
\omega (r,\theta , t=0)  = & \Omega_{\mbox{\scriptsize{surf}}} - \Omega_0
       \qquad  &\mbox{at } r = r_{\mbox{\scriptsize{in}}} \nonumber\\
\omega (r,\theta , t=0)  = & 0  \hspace{1.8cm} 
        &\mbox{at }  r > r_{\mbox{\scriptsize{in}}}
\end{eqnarray}

\subsection{Viscosity suppression}

Active regions are thought to originate from the emergence of strong magnetic
flux loops, driven by the buoyant instability of toroidal flux tubes. This
emergence process is quite well investigated (see e.g. \citeNP{FMI:Freibg}),
and a comparison of theoretical calculations with observations has allowed to
draw the conclusion that large AR originate from the rise of toroidal flux
tubes with field strengths of $\sim 10^5\,$G and magnetic flux $\sim
10^{22}\,$Mx, lying at the bottom of the convective zone. Observational and
theoretical evidence alike suggest that a few tens of megameters below the
surface, the rising loop is fragmented into a tree-like structure, leading to
the observed size spectrum of magnetic concentrations in AR from large spots
through pores and knots down to facular points. The probable cause of the
fragmentation is that the field strength in the top of the loop is reduced
below the local equipartition field strength, allowing external turbulence to
penetrate the tube and fragment it by the mechanism of flux expulsion. 

The origin of ephemeral active regions (EAR) is less clear, although this is an
important issue for our present purposes, as EAR contribute to the observed
photospheric magnetic flux density by an amount comparable to large AR. Their
properties suggest that EAR may be the result of the buoyant instability of 
thinner and weaker toroidal flux tubes, situated at shallower depths in the
convective zone. One possibility is that this shallower and more fragmented 
toroidal field, in turn, results from the ``explosion'' of large emerging flux
loops with lower initial field strengths and fluxes than those causing large AR
(\citeNP{FMI+:explosion}). While this is not the only possibility, tentatively
accepting it simplifies our present problem, as in this case the subsurface
structure corresponding to EAR collectively can also be thought of as a
collection of magnetic trees, just perhaps with somewhat deeper and more
extended ``crowns'' than in the case of large AR.

\begin{figure}[!t]
\centering
\noindent
\begin{minipage}{55mm}
\resizebox{55mm}{!}{
\includegraphics{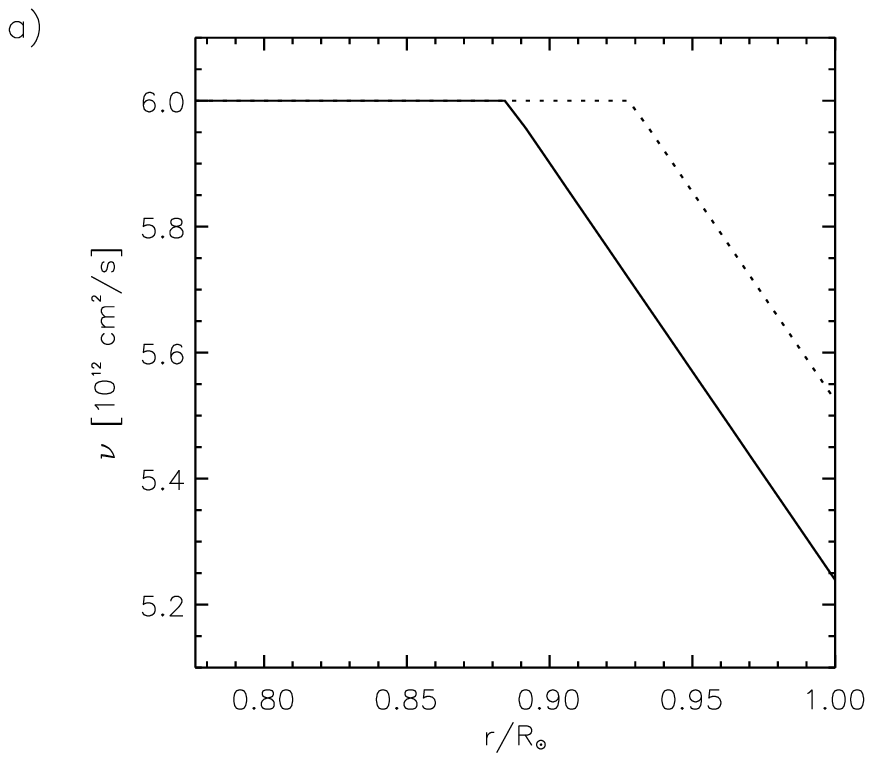}}
\end{minipage}
\begin{minipage}{55mm}
\resizebox{55mm}{!}{
\includegraphics{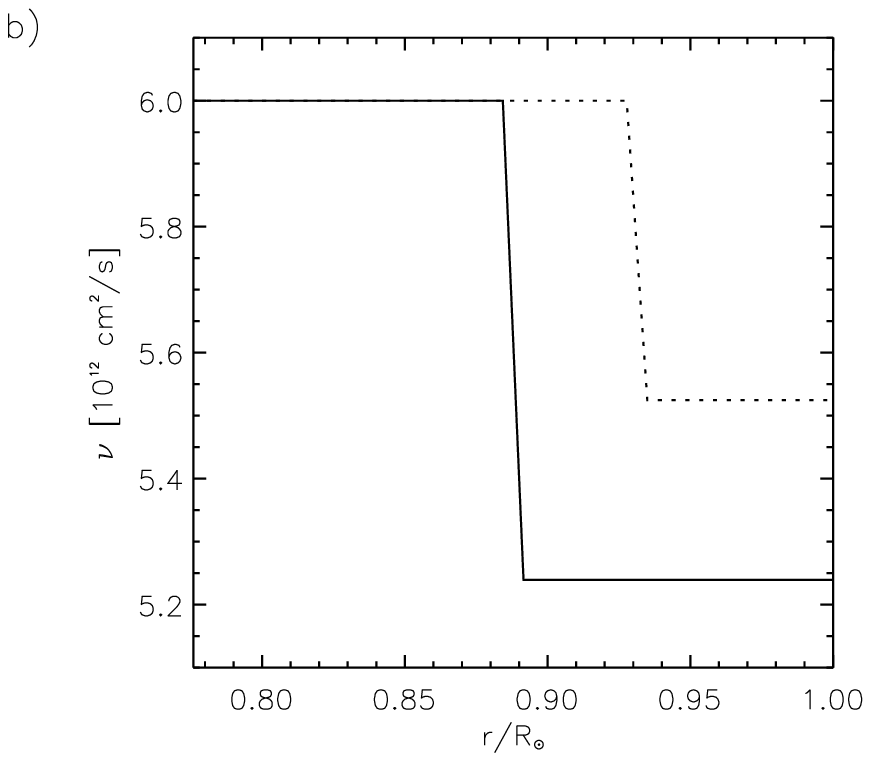}}
\end{minipage}
\caption{Radial profiles of the viscosity using the values $N_{AR}=3$, 
$\theta=60^{\circ}$ in equation (\ref{eq:nu}). Solid:
$D_0=z_0=80\,$Mm. Dashed: $D_0=z_0=50\,$Mm}
\label{fig:visc}
\end{figure}

The above considerations suggest that a simplified description of the 
subsurface magnetic structures that give the bulk of the flux density in the
upper convective zone may be a collection of magnetic trees, of crown
diameter $D_0$ and crown depth  $z_0$. Such a ``tree crown'' is essentially a
three-dimensional network of some characteristic scale $\lambda$. For those 
Fourier components of a turbulent flow whose spatial scale exceeds $\lambda$,
such a network will effectively present an impermeable ``wall'', thus these
modes will not contribute to turbulent transport. Assuming a Kolmogorov
spectrum, the turbulent diffusivities (including viscosity) will then be
reduced by a factor $(l/\lambda)^{4/3}$, $l$ being the integral scale of
turbulence. For $l\gg\lambda$ (which is the case expected, except for very
shallow depths) this reduction is so strong that one can basically assume
$\nu=0$ inside the ``crowns'' of magnetic trees. This ``high-wavenumber
filtering'' effect of AR magnetic structures on turbulence is confirmed by the
observations of abnormal granulation in plage areas
(e.g.\ \citeNP{Sobotka+:abnorm.granu}).

As our model is axisymmetric, the effective viscosity $\nu$ to be substituted
in our equations is in fact an azimuthal average. Accepting
$\nu_0=6\times10^{12}$ cm$^2$/s for the unperturbed value of the viscosity (a
value based on Babcock--Leighton models of the solar cycle),
then the relative vicosity perturbation $\nu'/\nu_0\equiv  1-\nu/\nu_0$ is
simply the expected value of the fraction of an azimuthal circle that passes
through the ``tree crowns''. A simple geometrical consideration shows that this
fraction is
\begin{equation}
  \frac{\nu'}{\nu_0}\equiv  1-\frac{\nu}{\nu_0}= 
  \frac{2 N_{AR} D_0}{2 \pi R_{\odot} \sin\theta}\, A(\theta,t) f(r),
  \label{eq:nu}
\end{equation}
where $N_{AR}$ is the typical number of active regions at maximum activity,
$f(r)$ is a radial profile, and $A(\theta,t)$ describes the time-colatitude
distribution (butterfly diagram) of AR. For $A(\theta,t)$ we use the same
expression as \citeN{Petrovay+Szakaly:2d.pol}:
\begin{eqnarray}
  \lefteqn{A=F(\theta,t;A_1,\Gamma_1,\Lambda_1,\delta_1)
  +F(\theta,t;A_2,\Gamma_2,\Lambda_2,\delta_2)}\\
  \nonumber \\
  \lefteqn{ F(\theta,t;A_i,\Gamma_i,\Lambda_i,\delta_i) = } \\
  && A_i \left[ 1+ \exp (\Gamma_i (\pi/4-\theta)) \right] ^{-1} 
  \cos \left[ Pt + 2\pi/\Lambda_i(\pi/2-\theta) + \delta_i \right] . \nonumber
\end{eqnarray}

The radial profile $f(r)$ must clearly fulfil the conditions $f(r=R_\odot)=1$
and $f(r\le {R_\odot-z_0})=0$. In the regime ${R_\odot-z_0}<r<R_\odot$ we use
two alternative profiles: triangular and rectangular, as illustrated in 
Fig.~\ref{fig:visc}. For $z_0$, two values are considered (50 and 80 Mm); 
$z_0=D_0$ is assumed throughout this paper.

\subsection{Numerical method}

We used a time relaxation method with a finite difference scheme first order
accurate in time to solve the equations. A uniformly spaced grid, with spacings
$\Delta r$ and  $\Delta \theta$ is set up with equal numbers of points in the
$r$ and $\theta$ directions. $r$ is chosen to vary from 
$r_{\mbox{\scriptsize{in}}}$ to $r_{\mbox{\scriptsize{surf}}}$ as given above, 
and $\theta$ varies from $0$ to $\pi/2$. The system is allowed to evolve until
it relaxes to a very nearly periodic behaviour (after about 100 cycles).

Our calculations are based on a more recent version of the solar model of 
\citeN{Guenther}.

\begin{figure}[!t]
\centering
\includegraphics[width=11.5cm]{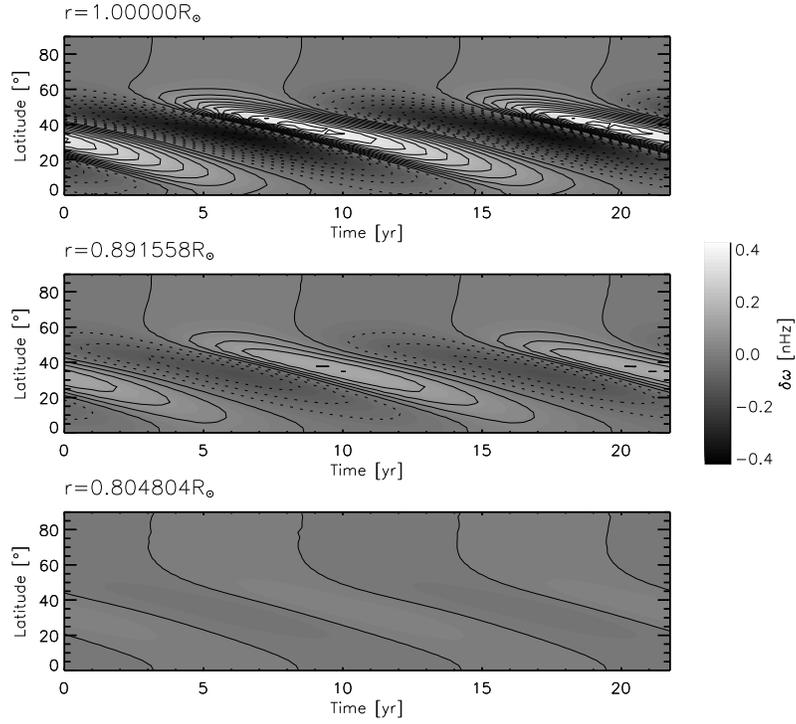}
\caption{Variation of the residual rotation rate $\delta\omega$ with latitude 
and time at a few selected radii as marked above each panel. In this
calculation we used the radial viscosity profile shown in
Fig.~\ref{fig:visc}(a), with $D_0=z_0=80\,$Mm,  for an equatorward propagating
wave model ($\Gamma_1=16,\ \Lambda_1=\pi/2,\ \delta_1=0,\ A_2=0$), and the
meridional circulation is neglected. The contours are drawn at intervals of
$0.04\,$nHz; dashed contours correspond to negative values.
}
\label{fig:butterfly_a}
\end{figure}

\begin{figure}[!h]
\centering
\includegraphics[width=12.5cm]{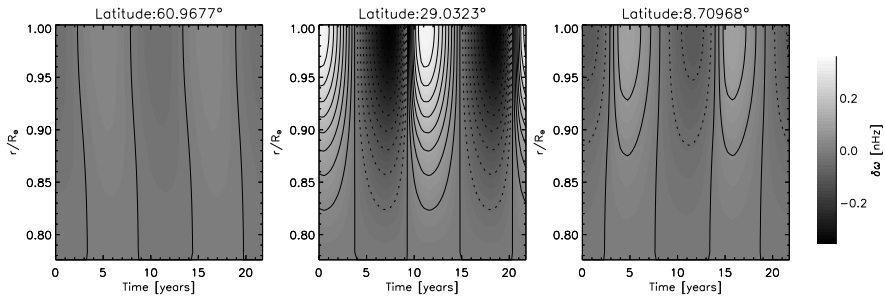}
\caption{Radial contours of the angular velocity residuals $\delta\omega$ as a
function of time at a few selected latitudes for the case shown in
Fig.~\ref{fig:butterfly_a}. The contours are drawn at intervals of $0.04\,$nHz.}
\label{fig:radius_a}
\end{figure}

\begin{figure}[!t]
\centering
\includegraphics[width=12cm]{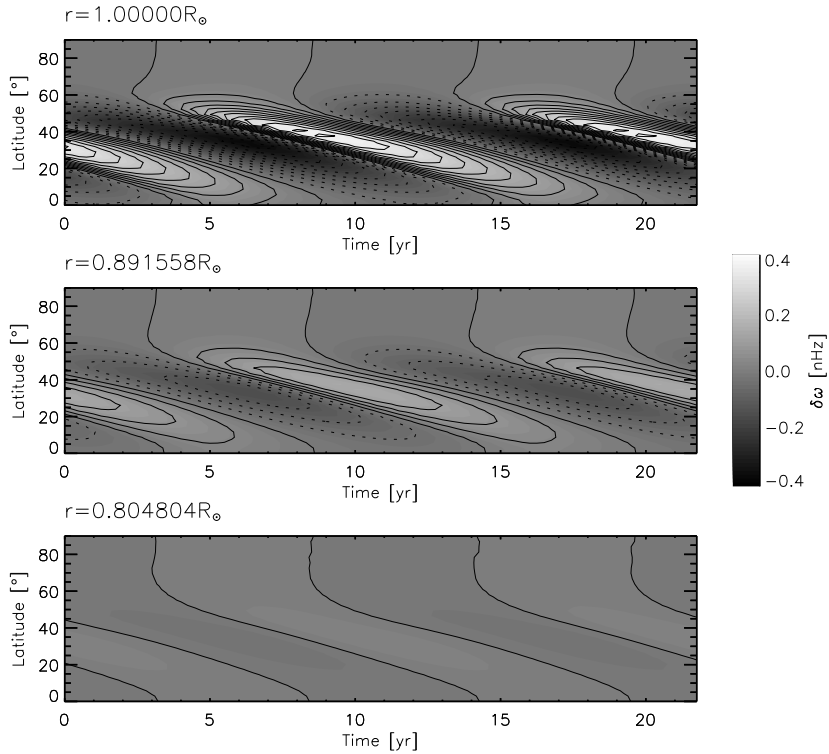}
\caption{Same as Fig.~\ref{fig:butterfly_a}, 
except that the meridional circulation is {\it{not}} neglected.}
\label{fig:butterfly_f}
\end{figure}

\begin{figure}[!h]
\centering
\includegraphics[width=12.5cm]{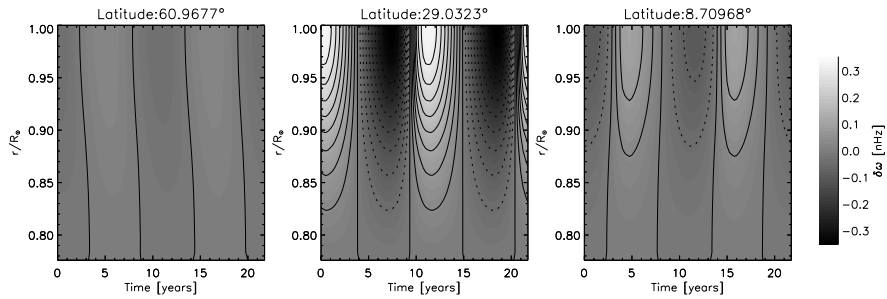}
\caption{Same as Fig.~\ref{fig:radius_a}, but the meridional circulation is 
{\it{not}} neglected.}
\label{fig:radius_f}
\end{figure}

\begin{figure}[!t]
\centering
\noindent
\begin{minipage}{55mm}
\resizebox{60mm}{!}{
\includegraphics{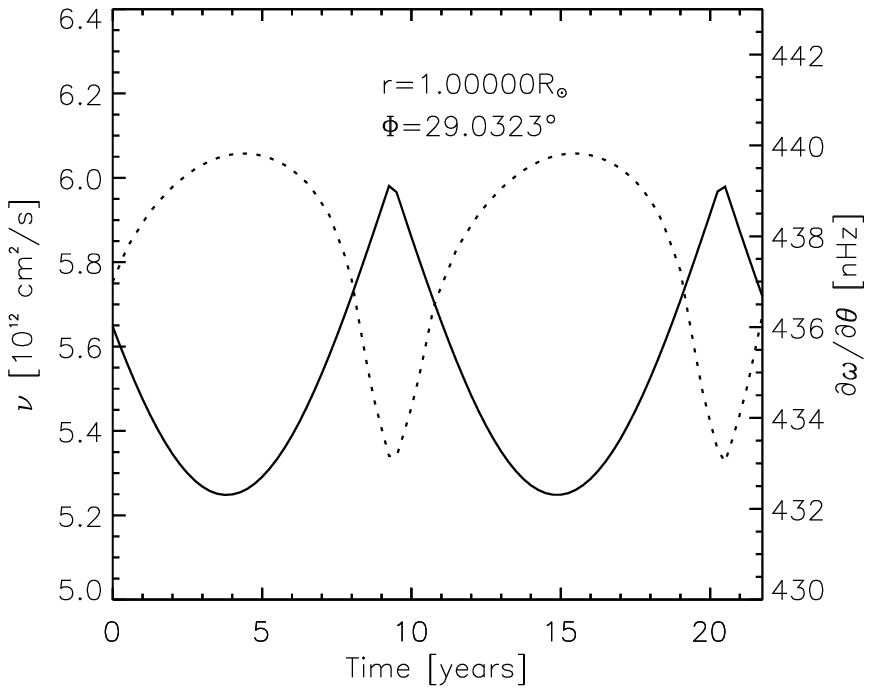}}
\end{minipage}
\begin{minipage}{55mm}
\resizebox{60mm}{!}{
\includegraphics{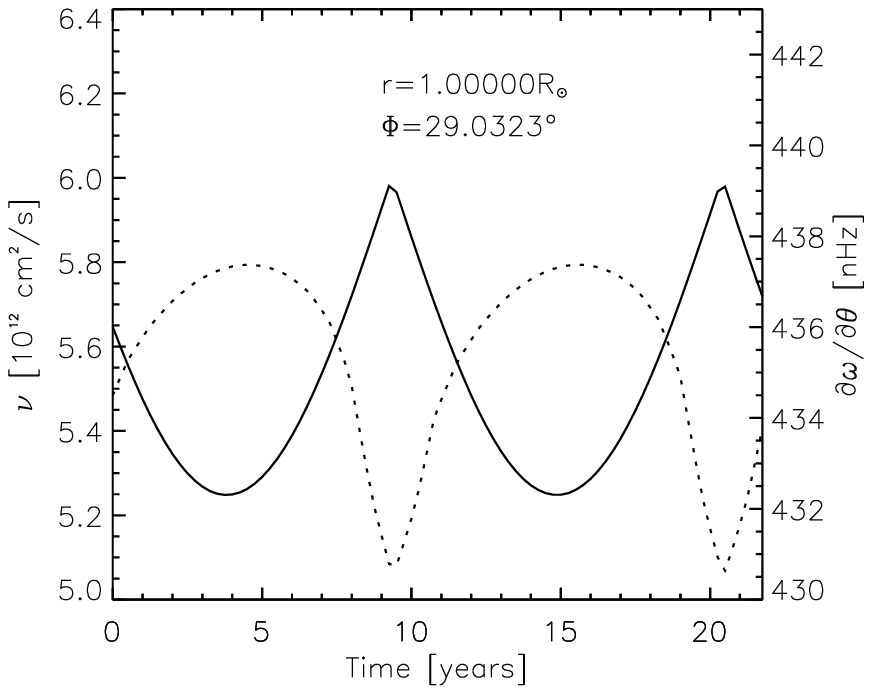}}
\end{minipage}
\caption{Variations of the viscosity (solid) and $\partial_{\theta}\omega$
(dashed) with time. {\it Left-hand panel:} the case 
where the meridional circulation is neglected (Fig.~\ref{fig:butterfly_a}).
{\it Right-hand panel:} the case where the meridional 
circulation is {\it{not}} neglected (Fig.~\ref{fig:butterfly_f}).}
\label{fig:domegadth_nu_time}
\end{figure}



\section{Results}


In order to reveal the migrating banded zonal flows, in the following we will
plot the distributions of the {\it residual\/} rotation rate, subtracting a
temporal average:
\begin{equation}
\delta\omega=\omega-\ov{\omega}
\end{equation}
At the bottom of our shell we clearly have $\delta\omega=0$.

First we present the results of a calculation neglecting the meridional flow 
($M=C=0$ in equation (\ref{eq:main})),
and using a butterfly diagram with an equatorward branch only  ($A_2=0$;
$\Lambda_1>0$). Here we used the radial viscosity profile presented in
Fig.~\ref{fig:visc}(a), with $D_0=z_0=80\,$Mm. Figure~\ref{fig:butterfly_a}
shows the evolution of the residual rotation rate $\omega$ as a function of 
latitude at three different depths. 

In order to study the radial behaviour of the zonal flow, we have plotted
$\delta\omega$ in Figure \ref{fig:radius_a} as a function of radius and time at a few
selected latitudes. It can be seen that the zonal flow pattern penetrates into
the convection zone, but the depth of the incursion is only about $0.1
R_{\odot}$.

Next, in Figs.~\ref{fig:butterfly_f}--\ref{fig:radius_f} we 
consider the effect of switching on the meridional circulation. 
It is found that the circulation has only a very weak effect on
the zonal flow pattern. 

Figure~\ref{fig:domegadth_nu_time} present
the variations of the viscosity and $\partial_{\theta}\omega$  with time. There
is clearly a near phase locking between the shear and the viscosity suppression
due to AR.

\begin{figure}[!t]
\centering
\noindent
\begin{minipage}{55mm}
\resizebox{55mm}{!}{
\includegraphics{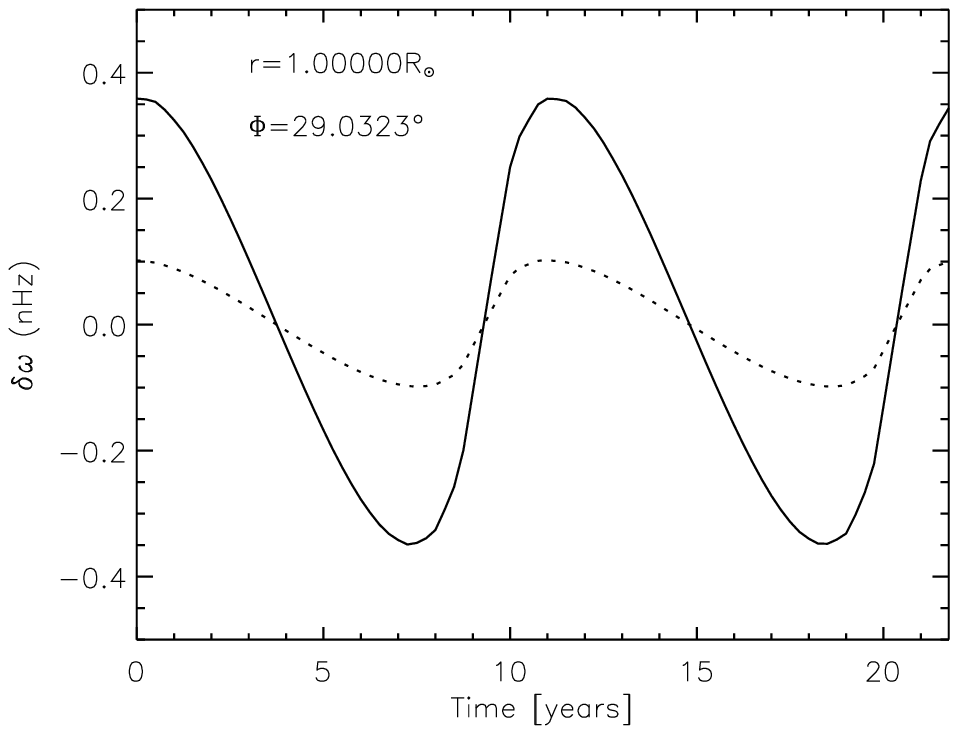}}
\end{minipage}
\begin{minipage}{55mm}
\resizebox{55mm}{!}{
\includegraphics{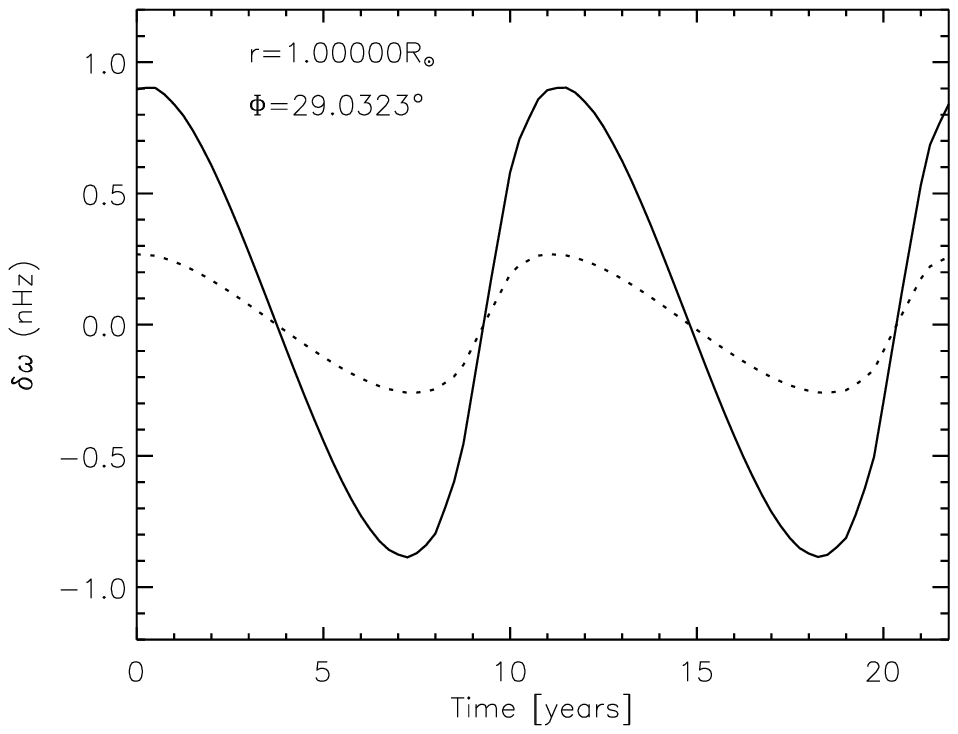}}
\end{minipage}
\caption{Amplitude of the residual rotation rate at the surface, at a
latitude of $30^{\circ}$ as a function of time. The solid line  corresponds to 
$D_0=z_0=80\,$Mm and the dashed line to $D_0=z_0=50\,$Mm.
{\it Left-hand panel:} using triangular viscosity profile, 
Fig.~\ref{fig:visc}a.
{\it Right-hand panel:} using rectangular viscosity profile, 
Fig.~\ref{fig:visc}b.}
\label{fig:amplitude}
\end{figure}



\begin{figure}[!t]
\centering
\includegraphics[width=12cm]{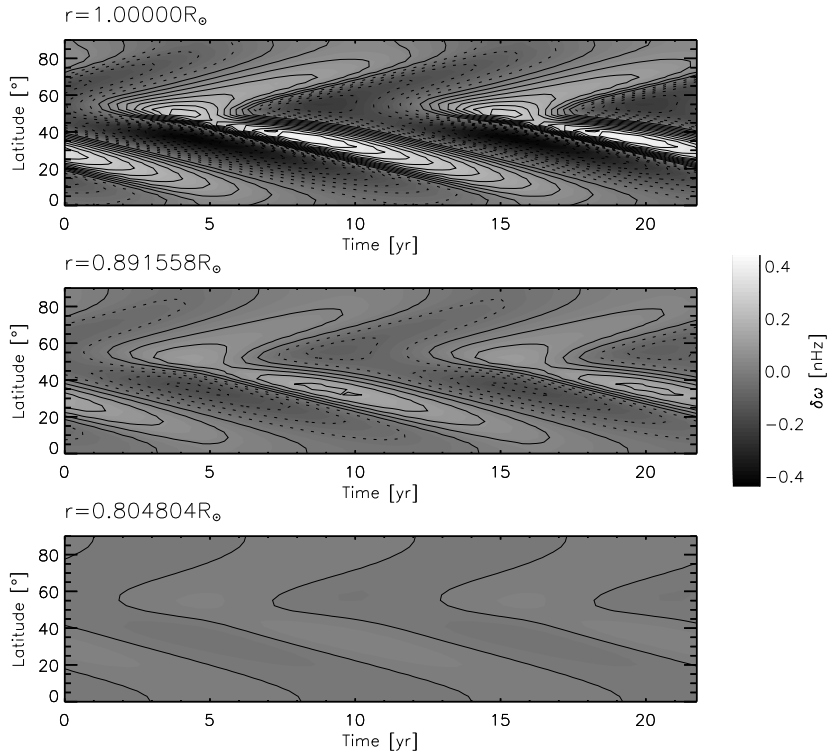}
\caption{Variation of the residual in rotation rate with latitude and time for
a migrating double wave model  ($A_1/A_2=4$, $\Gamma_1=-\Gamma_2=16$,
$\Lambda_1=-\Lambda_2=4\pi/9$, $\delta_1=0$, $\delta_2=\pi/4$).}
\label{fig:butterfly_h}
\end{figure}

\begin{figure}[!h]
\centering
\includegraphics[width=8cm]{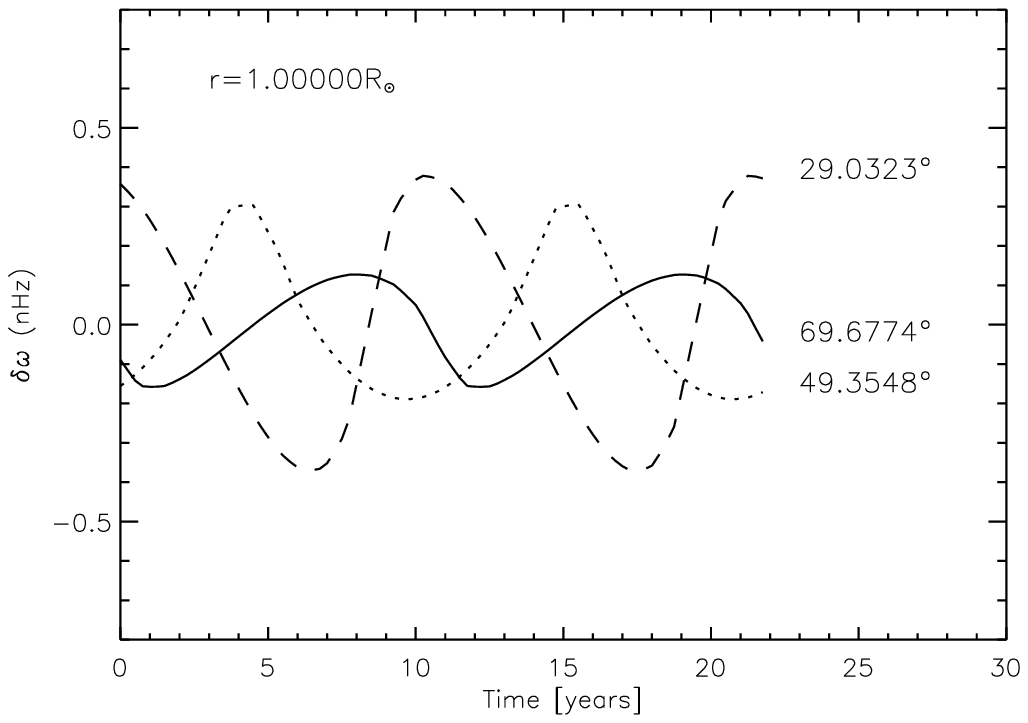}
\caption{Amplitude of the residual in rotation rate at different 
latitudes on the surface, as a function of time.}
\label{fig:amplitude_h}
\end{figure}

In order to study the influence of the choice of radial profile of the
viscosity on the amplitude of the oscillation, we have plotted $\delta\omega$
as a function of time using the different radial  profiles of the viscosity in
our calculations. (Figure~\ref{fig:amplitude}). Unsurprisingly, a  smaller
magnetic tree and/or a faster reduction of the viscosity suppression with depth
results in a smaller amplitude of the torsional  oscillation. The resulting
amplitudes are in general in the range $0.1$--$1\,$nHz, the highest values
being comparable to, though still significantly lower than the observed 
amplitudes. It is the amplitude that shows the greatest sensibility to our
model parameters, while the overall shape of the oscillation pattern and its
phase relation to the butterfly diagram does not change significantly with the
form of the viscosity suppression used.

Finally, Figs.~\ref{fig:butterfly_h}--\ref{fig:amplitude_h} show the 
results for a case where a full, double-wave butterfly diagram is used 
(\citeNP{Makarov+Sivaraman}). The polar branch, traced by polar faculae, is
assumed to be due exclusively to EAR in the present context. 

\section{Conclusion}

We have calculated the spatiotemporal variations of the solar rotation rate in
the upper convective zone caused by the suppression of turbulent viscosity due
to the presence of magnetic structures associated with (both large and
ephemeral) active regions. For concreteness we assumed that these structures can
be represented by an assembly of ``magnetic trees'' with ``crown heights''
$z_0\sim 50$--$80\,$Mm, and that turbulent transport inside these 
``tree crowns'' is effectively absent. 

Beside being a simplification, this whole scenario is clearly somewhat
speculative, especially as to the subsurface structures corresponding to
ephemeral active regions. Nevertheless we believe that this constitutes the
best ``educated guess'' that can be given at present. Still, keeping in mind
the uncertainties in the underlying modelling assumptions and in the parameter
values of our model, it is important to distinguish between features that seem
to be robust and valid for all models of this type from those that are sensible
to model details. Apparently robust conclusions that are valid for all the
models studied are
\begin{lista}
\item The zonal flows penetrate the convective zone down to a depth of $\sim
z_0$ only, i.e. to about $0.1\,R_\odot$, as shown by observations.
\item There is a very good phase coherence between the shear
$\partial_\theta\omega$ and the viscosity suppression (and therefore the level
of activity). We think that this phase relationship is a very important test of
models for torsional oscillations. Banded, migrating zonal flows can be produced
by a wide variety of models, so such tests are needed to determine which models
are compatible with observations also in their details.
\item The migration pattern (``butterfly diagram'') of the zonal flows
approximately reflects that of the active regions. In particular, if polar
faculae are interpreted as a sign of scattered activity in the form of EAR, then
a polar branch also arises naturally in the zonal flow migration pattern, as
observed. Indeed, our Fig.~\ref{fig:butterfly_h} bears a marked resemblance to
Figure~3 in \citeN{Howe+:SOGO}.
\end{lista}

On the other hand, one feature of the models that seems to be rather sensitive
to details of the model (e.g.\ triangular vs.\ rectangular viscosity profiles)
and to parameter values (especially $z_0$) is the {\it amplitude\/} of the
torsional wave. For the cases studied here, the amplitude is in general
significantly smaller than the observed amplitude of $\sim 5$--$10\,$nHz.  This
is true even though for suitable parameter combinations the model can produce
amplitudes that can reach a significant fraction of the observed amplitude. For
$z_0$ values higher than 80\,Mm (not shown here) we found that our model can
reproduce the full observed amplitude, but at the price of a deeper radial
penetration of the zonal flow than observed. Given our present ignorance, one
cannot exclude a scenario wherein EAR fields extend down to such great depths, 
and the shallowness of the torsional oscillations is due to e.g.\ the
prevalence of equatorward angular momentum transport (e.g.\ $\Lambda$-effect)
in the layers below 70\,Mm.

Thus, while the overall properties, such as radial dependence and phase, of AR
generated torsional oscillations seem to suggest that this mechanism can 
be partly responsible for the generation of torsional oscillations, the results 
are not decisive yet.

Further theoretical and observational constraints on the nature of the 
subsurface magnetic structures associated with both large and ephemeral active
regions would be necessary for a better assessment the amplitude of AR-induced 
zonal flows. Other possible improvements of the present models include the
incorporation of this effect in a complete model of differential rotation and
angular momentum transport in the convective zone.

\acknowledgements 
We thank D. B. Guenther for making his solar model available. This work was 
funded by the OTKA under grants No.\ T032462 and T034998.


\begin{ao}
\\
K. Petrovay\\
E\"otv\"os University, Dept.~of Astronomy\\
Budapest, Pf.~32, H-1518 Hungary\\
E-mail: kris@astro.elte.hu
\end{ao}

\end{article}
\end{document}